\documentclass[10pt]{article}

\usepackage{amsmath}
\usepackage{amssymb}

\usepackage{graphicx}

\usepackage{cite}

\usepackage{color}

\usepackage[labelfont=bf,labelsep=period,justification=raggedright]{caption}

\makeatletter
\renewcommand{\@biblabel}[1]{\quad#1.}
\makeatother

\date{}

\begin{document}

\begin{flushleft}
{\Large \textbf{Hierarchicality of Trade Flow Networks Reveals
Complexity of Products} }
\\
Peiteng Shi$^{1}$, Jiang Zhang$^{1,\ast}$, Bo Yang$^{2}$, and
Jingfei Luo$^{1}$,
\\
\bf{1} School of Systems Science, Beijing Normal University,
Beijing, China
\\
\bf{2} Ministry of Commerce of the People's Republic of
China\footnote{This paper only represents the personal views of the
author}
\\
$\ast$ E-mail: zhangjiang@bnu.edu.cn
\end{flushleft}

\section*{Abstract}
With globalization, countries are more connected than before by
trading flows, which currently amount to at least 36 trillion
dollars. Interestingly, approximately 30-60 percent of global
exports consist of intermediate products. Therefore, the trade flow
network of a particular product with high added values can be
regarded as a value chain. The problem is weather we can
discriminate between these products based on their unique flow
network structure. This paper applies the flow analysis method
developed in ecology to 638 trading flow networks of different
products. We claim that the allometric scaling exponent $\eta$ can
be used to characterize the degree of hierarchicality of a flow
network, i.e., whether the trading products flow on long
hierarchical chains. Then, the flow networks of products with higher
added values and complexity, such as machinery\&transport equipment
with larger exponents, are highlighted. These higher values indicate
that their trade flow networks are more hierarchical. As a result,
without extra data such as global input-output table, we can
identify the product categories with higher complexity and the
relative importance of a country in the global value chain solely by
the trading network.


\section*{Introduction}
As the process of globalization accelerates, countries throughout
the world are more connected, and collaboration is proceeding in an
unprecedented manner under the background of integrated global
markets of capital, labor force and products. Consequently, some
cross-border production chains, which comprise several countries or
regions, have inevitably emerged as the result of international
labor force division and collaboration at the global level
\cite{UNCTAD_global_2013,Gibbon_Governing_2008,Gereffi_governance_2005}.
However, because of the heterogeneities of products, the production
networks are very inhomogeneous. Some products in the electronics
and automotive industries, such as PCs or automobiles, can be broken
down into several independent components and easily transported and
assembled in different countries\cite{UNCTAD_global_2013}.
Therefore, a large fraction of imports for these products are not
for final consumption but, rather, are for re-production with higher
value-added and
exports\cite{Kotha_managing_2013,Rainnie_global_2013,UNCTAD_global_2013}.
Conversely, the networks for agriculture or raw material products
may have much shorter production chains. Thereafter the major
imports of these products are for final consumption.

Differentiating these products according to the length of their
production chains and the level of added-values is of importance for
countries' long-term development strategies. The conventional method
\cite{Tukker_global_2013,Lenzen_building_2013,Koopman_give_2010}
attempts to build the value flow networks among different products
directly by incorporating international input-output
tables\cite{leontief_input-output_1966,Miller_input-output_2009,Raa_economics_2005}.
Although the whole picture of production networks can be captured in
detail, obtaining accurate raw data on the global level is not
easy\cite{Koopman_give_2010,Miller_input-output_2009}. However, the
highly detailed international trade flow data for various products
among countries are well documented with a long
history\cite{feenstra_world_2005,zhu_compilation_2011}. In
particular, all bilateral trade flows are classified by different
products according to the SITC (Standard International Trade Coding)
or other equivalent coding methods. Therefore, a unique flow
structure of one product category can be extracted from the
international trade data.

In recent years, the world wide trade network as a specific instance
of a complex network has been studied
\cite{Hidalgo_product_2007,Hidalgo_building_2009,Fagiolo_world-trade_2009,garlaschelli_fitness-dependent_2004}.
Although both the common features shared by various complex networks
and some unique patterns are found, very little attention has been
paid to multi-networks of different
products\cite{Barigozzi_multinetwork_2010}. In this paper, we
attempt to discriminate products on their level of complexity and
value-added by identifying their unique trade flow structures. This
is possible because trade networks contain information on global
production networks. Almost all of the cross-border product flows in
the global value chain are recorded in the international trade data.

Our methodology is to compare the allometric scaling exponents among
the flow networks of different
products\cite{West_general_1997,Banavar_size_1999}. The allometric
scaling pattern has been found to be ubiquitous for trees spanned by
binary networks\cite{Banavar_size_1999,Garlaschelli_universal_2003},
such as food
webs\cite{Garlaschelli_universal_2003,zhang_scaling_2010}, trade
webs\cite{duan_universal_2007} and biological
networks\cite{Herrada_universal_2008}. Our previous work has
incorporated the flow analysis methods developed in ecology to
reveal the common nature of the flow networks in
general\cite{zhang_scaling_2010,zhang_allometry_2013}. It is natural
to extend this method to trade flows in which the allometric scaling
exponent is given a new explanation, the degree of hierarchicality.
This measure characterizes whether the product flows along a long
hierarchical chain. We calculate the allometric exponent of each
flow network in different product classifications and find that the
manufactured products with higher added values have larger
exponents. Furthermore, most exponents are larger than one,
indicating that the networks are hierarchical, whereas the networks
of the primary products with relative low added values have smaller
exponents, and the networks are flat. Hierarchicality always
indicates inequality and monopoly. We further calculate the relative
importance of each country in a product trading network and compare
the heterogeneities of the country's impact distribution for
different products using the GINI coefficient of country's impact.
Finally, the dynamics of allometric scaling exponents along time are
shown, and the globalization process can be interpreted.

\section*{Results}

\subsection*{Trade Flow Networks}
We use two datasets for study and comparison to eliminate the
potential discrepancies in the data. The fist one is from Feenstra
et al's ``World Trade Flows: 1962-2000'' dataset based on the United
Nations COMTRADE database (abbreviated as the UN
dataset)\cite{feenstra_world_2005}. This dataset covers the
bilateral trade flows of approximately 800 types of products
according to the SITC 4 (Standard International Trade Classification
system, Rev.4) classification standard from 1963 to 2000. In
addition, mainly the results from the year 2000 are shown and
discussed in the main text. Another dataset (the OECD dataset) is
the bilateral trade data from 2009 that was compiled by the
Organization of Economic Co-operation and Development
(OECD)\cite{zhu_compilation_2011}. The OECD dataset contains only
the OECD member countries, so the total number of countries is
smaller than the UN dataset. However, these countries dominate
approximately 70-80\% of the trade volume in the world. The products
classification standard of the OECD dataset is ISIC Rev. 3
(International Standard Industrial Classification of All Economic
Activities, Rev. 3), which is different from the SITC 4
classification. Please see the detailed discussions of the datasets
in the Supporting Information, SI.

The SITC4 codes are hierarchical, meaning that the categories with
longer codes are sub-categories of the ones with shorter codes if
they share the same prefix. For example, the product category 7 in
SITC4 is the category of machinery and transport equipment, so this
is a very generalized classification, whereas 71 and 72 are two
sub-categories of 7, representing the power machinery product and
vehicle categories, respectively.

\subsection*{Allometric Scaling of Trade Networks}
For each product trade network, we can define an exponent $\eta$ to
characterize the hierarchicality of the flow network. First, we need
to calculate two vertex-specific variables, namely, $T_i$ and $C_i$.

$T_i$, the trading volume of country $i$, is defined as the maximum
of $i$'s total imports or exports. This reflects the capacity of
trade flows through $i$. Next, $C_i$ is the impact of $i$ on the
entire network. It is defined as the total changes of trading volume
of other nodes on the network after the hypothetical deletion of
$i$. The concrete calculation of these two variables are referred to
the method section and supporting information.

\begin{figure}[!ht]
\begin{center}
\includegraphics[scale=0.7]{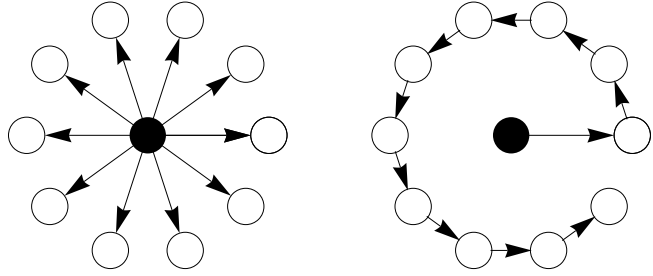}
\end{center}
\caption{Two special spanning trees with the minimum allometric
exponent 1 (left, a star network) and maximum exponent 2 (right, a
directed chain)} \label{fig.starchain}
\end{figure}

Typically, for various empirical trade networks, $C_i$ and $T_i$
have a strong correlation which can be described by a power law:
\begin{equation}
\label{eqn.allometricscaling} C_i\sim T_i^{\eta},
\end{equation}
where $\eta$ is the allometric scaling exponent. This equation is
extended from the empirical allometries from river basins, vascular
networks and food
webs\cite{Banavar_size_1999,Garlaschelli_universal_2003}. Previous
studies on spanning trees indicate that the exponent $\eta$ can be
used to reflect the hierarchicality or flatness of a tree. For
example, two extreme cases of spanning trees are shown in Figure
\ref{fig.starchain}. The star network that has the smallest
exponent, $1$, is the flattest tree, whereas the chain network that
has the largest exponent, $2$, is the most hierarchical tree.

This calculation can be extended to general flow
networks\cite{zhang_scaling_2010,zhang_allometry_2013}. Nevertheless
the exponent is not bound in $[1,2]$. However, we can also define
the exponent as the hiearchicality of a general flow network, as it
will contain long flow chains if its exponent is larger(see
Supplementary Information).

\begin{figure}[!ht]
\begin{center}
\includegraphics[scale=0.7]{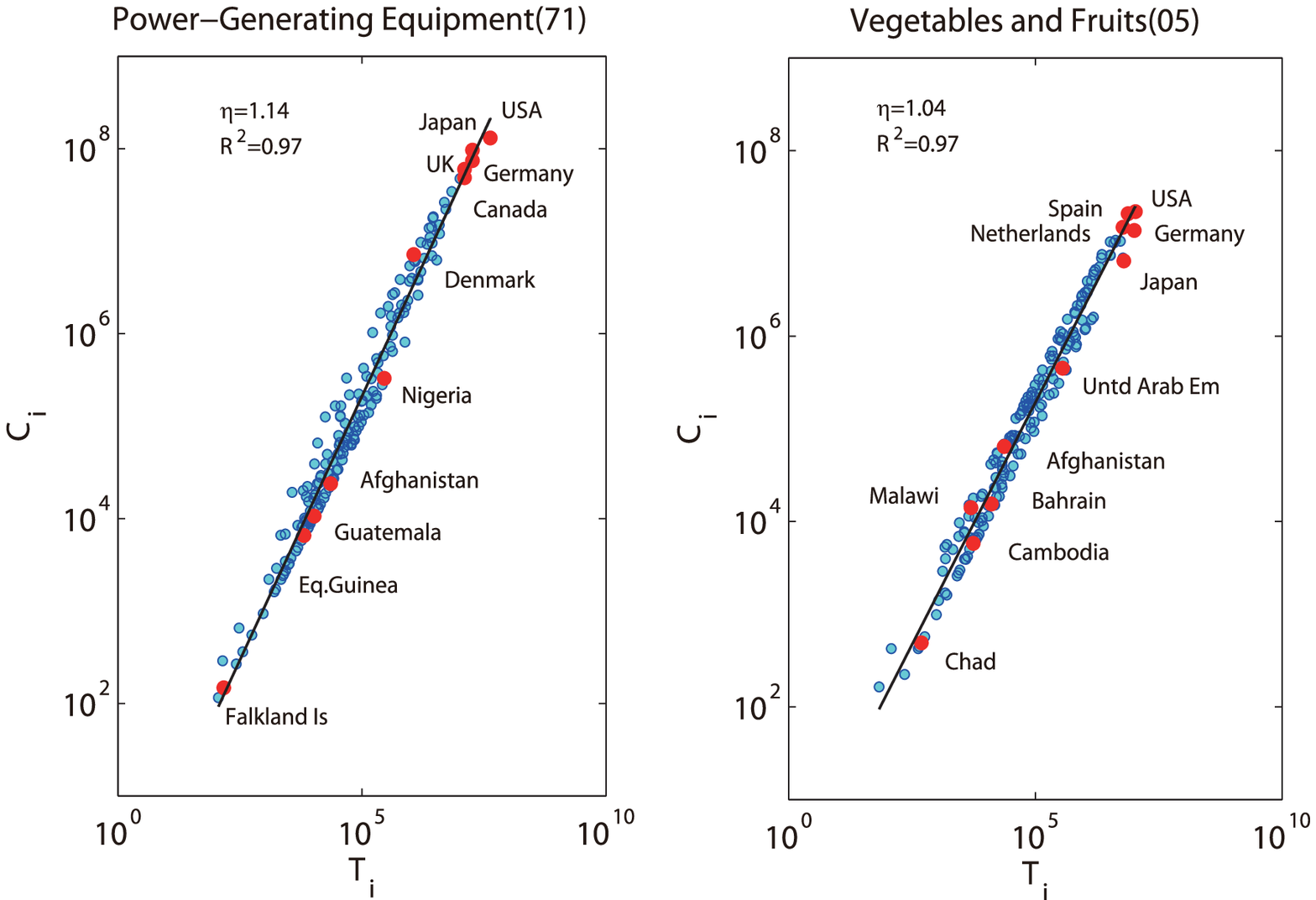}
\end{center}
\caption{The allometric scaling between $T_i$(in U.S. dollars) and
$C_i$(in U.S. dollars) of two networks. The left figure shows a
super-linear scaling law (with an exponent larger than 1) for power
generating products, whereas the right figure shows a nearly linear
scaling law (with an exponent close to 1) for fruits and vegetables}
\label{fig.allometricscaling}
\end{figure}

It turns out that the allometric scaling pattern (Equation
\ref{eqn.allometricscaling}) is very general for all of the studied
trade networks but their exponents are not similar. Figure
\ref{fig.allometricscaling} shows the allometric scaling patterns of
two products.

In Figure \ref{fig.allometricscaling}, each data point stands for a
country participating in the international trade of this product.
The pairs of $T_i$ and $C_i$ form a straight line on the log-log
coordinate, which means a power law relationship between the two
variables exists (i.e., Equation \ref{eqn.allometricscaling}). The
exponents for these two products are distinct, indicating that the
power-generating trade network is more hierarchical than the fruit
and vegetable networks. In other words, the production for power
generating machines is along a longer value-added chain than fruits
and vegetables.

\begin{figure}[!ht]
\begin{center}
\includegraphics[scale=0.8]{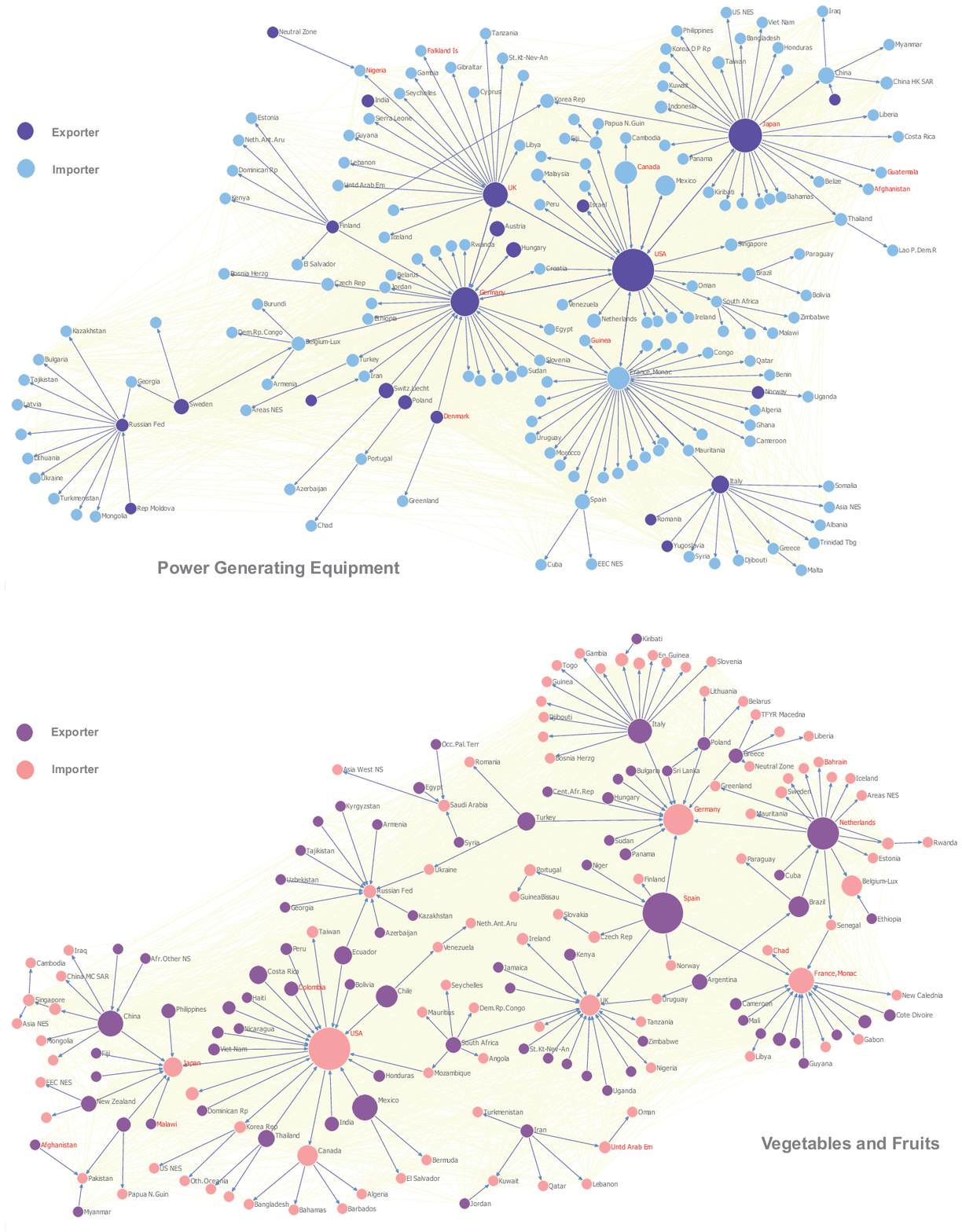}
\end{center}
\caption{Visualization of trade flow network for power generating
equipment (upper) and fruits and vegetables (lower). We use
different colors to distinguish nodes as importer (import is larger
than its export) and exporter (export is larger than import). The
size of node denotes the total volume of trade. In these two
networks, only the backbones are shown as the main parts and all
other un-important links are hidden as backgrounds. The backbone
extracting method is according to \cite{Foti_nonparametric_2011}}
\label{fig.visualization}
\end{figure}

This point can be visualized by the network plots of these two
products (Figure \ref{fig.visualization}). Although only the
backbone links are shown and other links are faded in the
background, it is clear that the upper network has many long chains
that always root from some major exporters of power generating
machines(e.g. the US and Japan). However, the lower network is more
fragmented. Although several large countries (e.g. the US) still
occupy a large fraction of the fruit trade, most of them are
importers. This implies the whole network lacks a center and is more
flat. Intuitively, that is the reason why the exponent of the first
network is larger than the latter.

\subsection*{Exponents Comparison and Distributions}

We further compare the exponents among different networks of
products in a more systematic way. In Tables \ref{tab.exponents} and
\ref{tab.exponents2}, we list exponents for all 1-digit products in
the UN dataset and OECD dataset to compare.

\begin{table}
\centering
 \caption{Exponents of 1-digit SITC4 categories in the UN dataset}
{\small (The categories of 8 (Miscellaneous) and 9(Not classified)
are ignored in this table, the last row shows the allometry of all
products as an integrated network.)}
 \label{tab.exponents}
\begin{tabular}{lllll}
\hline Code & Classification  & $\eta$ & $R^2$&GINI\\
\hline

7& Machinery and transport equipment&   1.136$\pm$0.026 & 0.974&0.889\\
6& Manufactured goods classified chiefly by materials& 1.120$\pm$0.026& 0.962&0.830\\
5& Chemicals and related products& 1.117$\pm$0.034& 0.972&0.877\\
1& Beverages and tobacco &  1.116$\pm$0.033 &0.958&0.868 \\
4& Animal and vegetable oils, fats and waxes& 1.077$\pm$0.029& 0.973&0.847\\
0& Food and live animals& 1.043$\pm$0.032& 0.971&0.798\\
3& Mineral fuels, lubricants and related materials&1.042$\pm$0.018&0.954&0.821\\
2& Crude materials, inedible, except fuels& 1.001$\pm$0.020& 0.988&0.815\\
-& All Products&1.022$\pm$0.030&0.965&0.817\\
 \hline
\end{tabular}
\end{table}

\begin{table}
\centering
 \caption{Exponents for different products in the OECD dataset}
 {(\small The products in different industries coded by the ISIC Rev. 3 coding system for industries is shown.
 The financial intermediation, business services, wholesale and retail trade, transport and storage, post and telecommunication, hotels and restaurants, and construction industries are ignored because their trades do not stand for goods flows.
 The last row shows the allometry of all industries as an integrated network)}
 \label{tab.exponents2}
\begin{tabular}{lllll}
\hline Code & Classification  & $\eta$ & $R^2$&GINI\\
\hline

29 & Machinery and equipment, nec&    1.146$\pm$0.072&     0.947&0.656\\
23T26& Chemicals and non-metallic mineral products& 1.129$\pm$0.079&0.937&0.563\\
34T35& Transport equipment& 1.124$\pm$0.075&0.941&0.669\\
30T33& Electrical and optical equipment& 1.112$\pm$0.070& 0.948&0.667\\
27T28& Basic metals and fabricated metal products& 1.092$\pm$0.080&0.974&0.568\\
40T41& Electricity, gas and water supply& 1.075$\pm$0.054& 0.931&0.649\\
36T37& Manufacturing nec; recycling& 1.074$\pm$0.078& 0.967&0.684\\
15T16& Food products, beverages and tobacco& 1.073$\pm$0.081& 0.931&0.553\\
20T22& Wood, paper, paper products, printing and publishing&
1.051$\pm$0.088& 0.926&0.589\\
10T14&   Mining and quarrying &   1.019$\pm$0.041&  0.911&0.721\\
01T05& Agriculture, hunting, forestry and fishing&
1.019$\pm$0.070&0.978&0.576\\
17T19&Textiles, textile products, leather and footwear& 0.998$\pm$0.065& 0.949&0.705\\
-&All industries&0.941$\pm$ 0.072&0.924&0.474\\
 \hline
\end{tabular}
\end{table}

Both tables display large gaps of exponents for different products
($[1.001,1.136]$ for the UN dataset and $[0.944,1.146]$ for the OECD
dataset). Although some slight differences between SITC4
classification and ISIC Rev. 3 classifications exist, the products
of machinery, equipment, chemicals and similar are of higher
exponents than the products of food, mining and agriculture
industries. This unique observation can be further confirmed and
extended to finer classifications.

\begin{figure}[!ht]
\begin{center}
\includegraphics[scale=0.6]{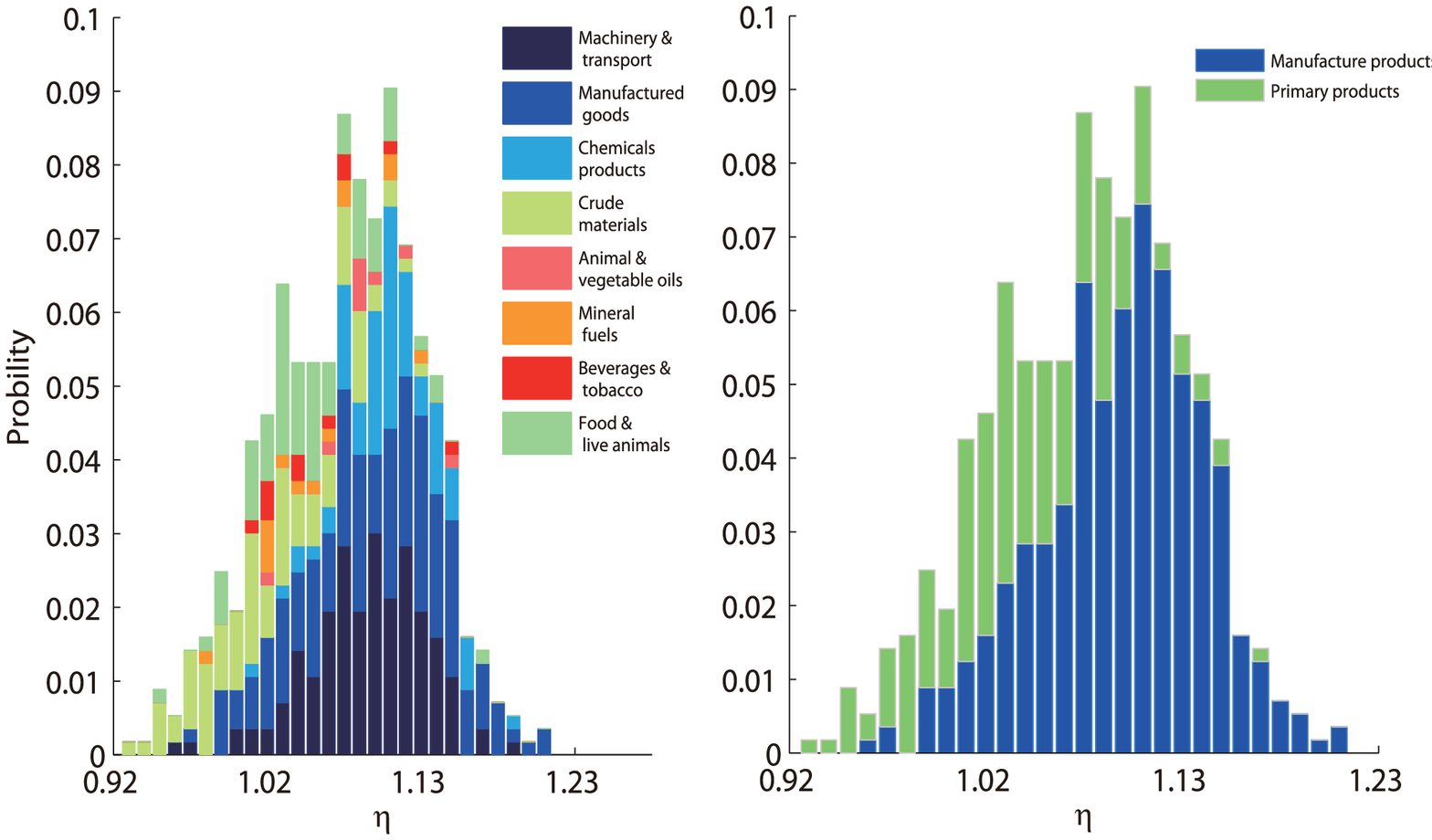}
\end{center}
\caption{Exponent Distribution for All of 4-digit SITC4 Product
Categories. The stacked bar charts of different colors correspond to
1-digit SITC4 categories (left) and primary and manufacture
classifications (right). For one specific 1-digit classification
(say 0 for food and living animals), we can calculate the
frequencies on each exponent intervals for all products with 0
prefix, then these frequencies as little bars are stacked on the
tops of existing bars.} \label{fig.exponentdistribution}
\end{figure}

Figure \ref{fig.exponentdistribution} shows the exponent
distribution of all products with 4-digit classification in the UN
dataset. The frequency curve has a bell-shape peaked at 1.09, which
means most product networks are hierarchical. The stacked color bars
show the distributions of all 1-digit classifications (Figure
\ref{fig.exponentdistribution} left). Note that most blue bars are
located in the right side of the bell-shaped curve, whereas the
green and yellow bars are located on the left side, indicating that
the machinery and manufactured products have larger exponents than
the food and beverage products. This phenomenon is better
illustrated in the right subplot of Figure
\ref{fig.exponentdistribution}, in which we simply classify the
products as primary products (prefix of 0,1,2,3, and 4) and
manufactured products (prefix of 5,6,7,8, and 9).

\subsection*{Allometric Exponent and Product Complexity}

According to the observations, we know that the allometric exponents
of the trade flow network can reflect the basic properties of
products. The manufactured products with higher added-value and
complex production process always have larger exponents. Therefore,
we conjecture that a positive correlation between the exponents and
the nature of products (complexity or value added) may exist.

To test our hypothesis, we perform two correlation analysis on both
datasets. For the UN dataset, we correlate the exponents with PRODY,
one of the measurements of product complexity. It is calculated as
the average comparative advantage weighted by the GDP per capita of
the exporters of this product\cite{Hausmann_what_2007}. It is
calculated as follows:
\begin{equation}
\label{eq.prody} PRODY(p)=\sum_c Y_cRCA(c,p),
\end{equation}
where, $Y_c$ is the GDP per capita of country $c$, and $RCA(c,p)$ is
the comparative advantage of country $c$ exporting $p$. The
summation is taken for all of the countries exporting $p$.
$RCA(c,p)$ can be calculated as
$RCA(c,p)=\frac{E(c,p)/\sum_pE(c,p)}{\sum_c{(E(c,p)/\sum_pE(c,p))}}$,
where $E(c,p)$ is the total export value of $c$ on $p$.

\begin{figure}[!ht]
\begin{center}
\includegraphics[scale=0.6]{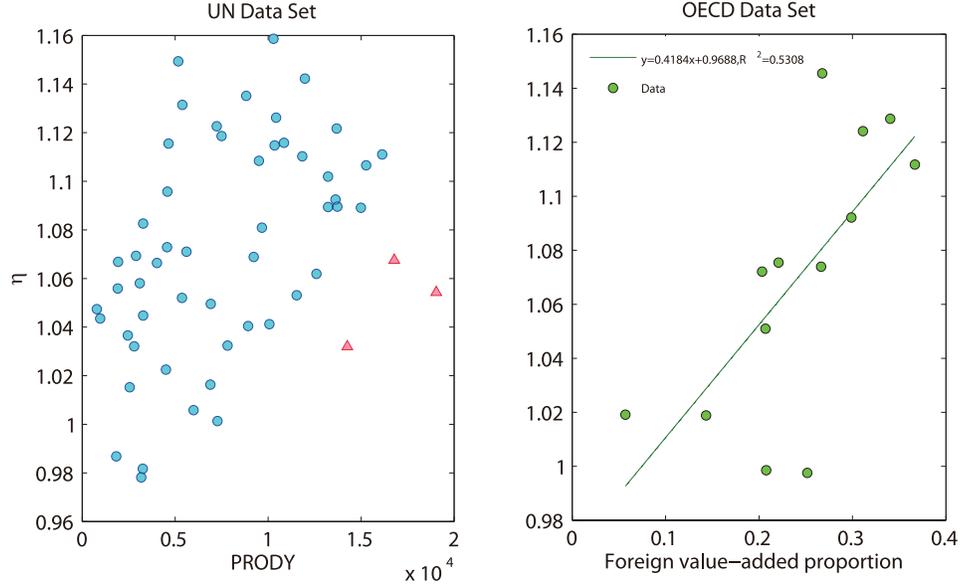}
\end{center}
\caption{The relationship between $\eta$ and PRODY of each 2-digit
classification(left) in the UN dataset and $\eta$ versus mean
proportion of foreign value added for products in the OECD dataset.}
\label{fig.correlation}
\end{figure}

Figure \ref{fig.correlation} shows the relationship between the
exponent $\eta$ and PRODY of each product using the 2-digit
classification of the UN dataset. The correlation coefficient of
these two variables is 0.37, and it can be improved to 0.44 if the
three outliers (triangles) in Figure \ref{fig.correlation} are
omitted.

For the OECD dataset, the domestic and foreign value added for each
product-country combinations are available (see SI). This enables us
to correlate exponents with the average foreign value-added ratio of
each product. Here, the proportion of foreign value added is the
ratio between the total value added and gross export for all of
countries exporting this product\cite{UNCTAD_global_2013}. The
relationship between $\eta$ and foreign value-added proportion is
shown in the right plot of Figure \ref{fig.correlation}. There is a
clear positive correlation between them, and the correlation
coefficient is $0.692$.

Consequently, we conclude that the allometric exponent $\eta$ of
each trade flow network can characterize the complexity and
value-added proportion of given product. When a product needs more
complex production processes, more countries must be involved to
form a long value chain, so that more value is added on the product.
All of these properties must be reflected in the flow structure of
the product trade network. That is the reason why the allometric
exponent $\eta$ can be distinct for different products.

\section*{Discussion}

\subsection*{Country Impacts}

In addition to the structural properties of the entire network, the
node positions in the global value chain are also of importance and
interests. In our study, $C_i$, the total impact of country $i$
toward the entire network, can be viewed as a vertex centrality
indicator because it measures the degree of the entire network is
influenced if node $i$ is removed. This understanding is in
accordance with the standard HEM (Hypothetical Extraction
Method)\cite{cella_input-output_1984,song_linkage_2006} in
input-output analysis once the trade flow networks are understood as
an input-output matrixes.

\begin{figure}[!ht]
\begin{center}
\includegraphics[scale=0.6]{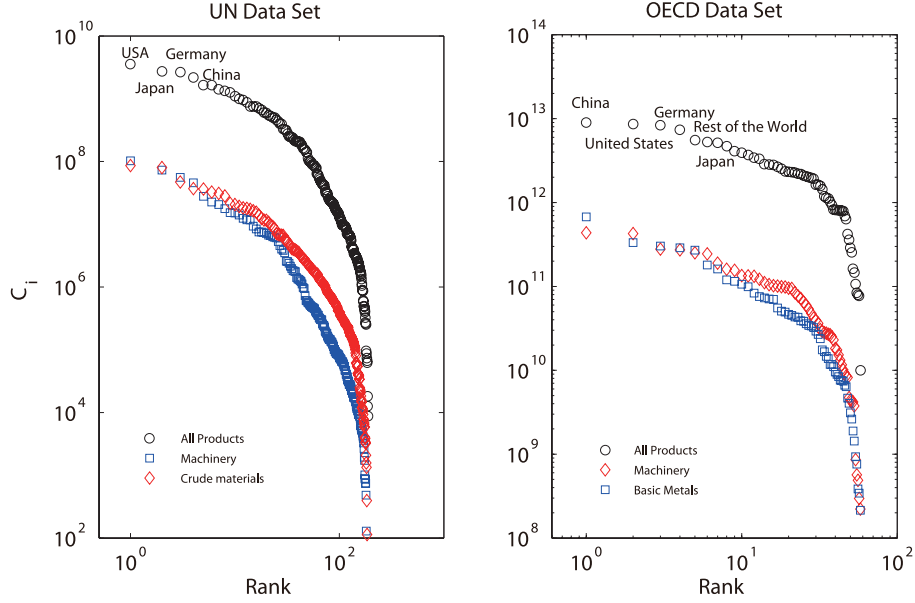}
\end{center}
\caption{$C_i$ Distributions of Both Datasets. The unit of $C_i$ is
the US dollar} \label{fig.cidistributions}
\end{figure}

Figure \ref{fig.cidistributions} shows the distributions of $C_i$
for trade networks of all products and several selected products
both in the UN and OECD datasets. In addition, the top 10 countries
are listed in the Supplementary Information.

\subsection*{Centrality and Inequality}

In our previous works exploring allometric scaling on ecological
flow networks\cite{zhang_allometry_2013}, the exponent $\eta$ is
explained as the degree of centrality, i.e., whether several large
nodes dominate and have a disproportional impact on the entire
network. This explanation can also be extended to this study. The
networks with higher $\eta$s are more centralized. Thus, a few large
countries can impact the entire network, in which the impact's
degrees $C_i$ are disproportional to their direct trade flow $T_i$.

For example, we have three flow networks with the same
$T_i=\{1,2,3,4,5\}$ but different $\eta=\{1,1/2,2\}$. Then, their
$C_i$s are $C_i^{(1)}=\{1,2,3,4,5\}$ for $\eta=1$,
$C_i^{(2)}=\{1,1.4,1.7,2,2.2\}$ for $\eta=1/2$ and
$C_i^{(3)}=\{1,4,9,16,25\}$ for $\eta=2$, respectively. As a result,
the largest country (the node with the largest $T_i$) dominates
$5/(1+2+3+4+5)\approx 33\%$, $2.2/(1+1.4+1.7+2+2.2)\approx 27\%$,
and $25/(1+4+9+16+25)\approx 45\%$ impacts of the entire networks,
respectively. Therefore, the third network is much more centralized
than the second network.

However, the inequality of exporting products is mainly from the
heterogeneity of the resource distribution but not the network
effect, which is characterized by $\eta$. For example, petroleum
export is heterogenous because of the unevenness of fossil fuel
resource distribution geographically. Therefore, new indicator is
needed.

We use the GINI coefficient of $C_i$ distribution to characterize
the overall inequality of the flow network structure. The $C_i$
distribution can account for both inequality origins: natural
resource distribution and network effects. First, it is obvious that
the natural inequality of resource distribution can be reflected by
the $T_i$ distribution. Suppose $T_i$ follows a Zipf law, $
T_i(r)\sim r^{-\alpha}$, where, $\alpha$ is the Zipf exponent, and
$r$ is the rank order of $i$. We know that there is a power law
relationship between $T_i$ and $C_i$ according to Equation
\ref{eqn.allometricscaling}. Thus, $C_i$ also follows the Zipf law:
$ C_i(r)\sim r^{-\beta}=r^{-\alpha\eta}$, where $\beta=\alpha\eta$
is its exponent. Therefore, the distribution of $C_i$ ($\beta$)
contains both types of information: natural heterogeneities
($\alpha$) and network effect($\eta$).

Although $C_i$ does not follow the Zipf distribution in our
empirical data (shown in Figure \ref{fig.cidistributions}), the
previous conclusion that the distribution of $C_i$ contains both
types of information, is still correct. Typically, the GINI
coefficient (bounded by [0,1]) can be used to characterize the
inequality of a variable no matter what type of distribution it
follows.

In the last column of Table \ref{tab.exponents}, we show the GINI
coefficients of all 1-digit product categories. Most products have
similar rank order by GINI as the order by $\eta$. However, the
order of manufactured goods (Code 6) falls down from No. 2 (by
$\eta$) to No. 5 (by GINI coefficient), and the order of food and
live animals falls from No. 6 to the bottom. These differences
indicate that these two types of products are not so unequal as
predicted by the exponent $\eta$ because the average trading volumes
($T_i$) distribute evenly among countries although their trading
networks are more centralized. In the last column of Table
\ref{tab.exponents2}, the GINI coefficients of all industries of the
OECD dataset are shown. There is a large deviation of the order by
$\eta$ from the GINI coefficients. Some industries, such as mining
and textiles, have high ranks of GINI coefficients but low ranks of
$\eta$. That indicates these industries are resource monopolized.
Basic metals and chemicals have high ranks of $\eta$ but low ranks
of GINI coefficients which means the trade networks of these
products are centralized.

Another interesting finding is the exponent of the integral trade
network that consists of all trading products is $1.02$ (It is
$0.94$ in the OECD dataset). This value is less than the mean
exponent when averaging all individual products. This is also found
for the GINI coefficients. These findings imply international trade
of all products in general becomes much more decentralized than each
single product's trade. Therefore, trade in diverse types of
products can make the world flatter. Though we still do not know to
what degree this conclusion could be true. This will be left for
further investigation.

\subsection*{Exponents in Different Years}

\begin{figure}[!ht]
\begin{center}
\includegraphics[scale=0.6]{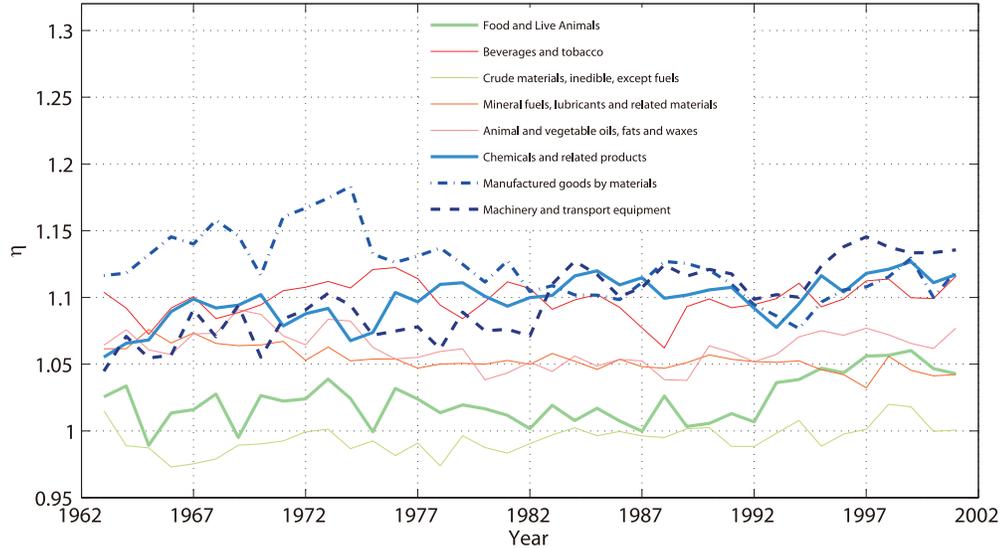}
\end{center}
\caption{Allometric exponents $\eta$s of 1-digit classification
products change with time } \label{fig.time}
\end{figure}

The UN dataset records the international trade data historically
from year 1962 to 2000. This enables us to study the dynamics of
exponents. In Figure \ref{fig.time}, we show how these exponents
change with time.

Most exponents are almost stable. However, machinery, transport
equipment and manufactured material goods have large changes. The
latter has very large exponents before 1982, but the former climbs
to the top 1 after approximately 1982. Note that some cross-border
companies emerged in and around the 1980s. Therefore, machinery and
transport equipment products, which depend on vertical labor
division but not material, has the largest exponent, whereas the
manufactured goods, which are more independent of global
cooperation, change in an opposite direction. Thus, the dynamics of
the exponents may reflect the globalization process.

\section*{Methods}

\subsection*{Flow Network Model}

A flow network model can be built for each product category. Nodes
on the network are countries. Directed edges are trading
relationships between countries and weights on edges are trading
flows measured by the unified money units (the US dollar in our
datasets).

If there are in total $N$ countries participating in trade of the
focus product $p$, then a flow network can be represented by an
$N\times N$ flux matrix $F^p$, in which the element $f_{ij}^p$
stands for the trade flow of $p$ from $i$ to $j$. The superscript
$p$ will be omitted to facilitate our expression. All of the
variables as well as the trade networks in the following sections
are defined for one specific product.

\subsection*{Trading Volume and Impact}

After the construction of the network, we can calculate two
important variables for each country. First, $T_i$ defined as the
trading volume of country $i$, is the maximum value of either import
or export,
\begin{equation}
\label{eq.ti} T_i=\max(\sum_{j=1}^{N}f_{ji},\sum_{j=1}^{N}f_{ij}).
\end{equation}

This value measures the amount of product $p$ flows through country
$i$, and $T_i$ reflects the flow capacity that country $i$ can
import or export $p$.

However, this indicator can only characterize the direct importance
of country $i$ without the network effect consideration. As trade
networks are understood as production chains in our study, one
country's export not only influences the direct neighbors but also
indirect neighbors along the production chain.

Therefore, we use another variable $C_i$ to indicate the impact of
country $i$ on the entire network. $C_i$ is defined as the total
reduction of trade volume of all countries if $i$ is deleted in the
network. We will introduce its calculation.

Before $C_i$ is defined, we should introduce another important
matrix $M$. It is analogous to the technical coefficient matrix in
input-output theory:
\begin{equation}
\label{eq.mmatrix} m_{ij}=\frac{f_{ij}}{T_{i}}.
\end{equation}
Thus, $m_{ij}$ measures the ratio of the export from $i$ to $j$ to
the total trade volume of $i$. Then, the following identity can be
derived:
\begin{equation}
\label{eq.mmmatrix} T=MT+S,
\end{equation}
where, $T=(T_1,T_2,\cdot\cdot\cdot,T_N)^T$,
$S=(S_1,S_2,\cdot\cdot\cdot,S_N)^T$. In addition,
\begin{equation}
S_i=T_i-\sum_{j=1}^N f_{ji}
\end{equation}
can be viewed as the total domestic value added from $i$ (see the
discussion in SI). Then, we can obtain an important identity from
Equation \ref{eq.mmatrix}:
\begin{equation}
\label{eq.important} T=(I-M)^{-1}\cdot S,
\end{equation}
where $I$ is the identity matrix. Now, suppose node $i$ is deleted
in the network. Then the $i$th column in $M$, and also $S_i$ will be
set to $0$ according to the HEM
method\cite{cella_input-output_1984,song_linkage_2006}. Suppose $M$
turns into $M'$ and $S$ turns into $S'$. Then, the new total trade
volume vector can be computed if we believe the identity Equation
\ref{eq.important} also holds for $M',S'$ and $T'$:
\begin{equation}
\label{eq.importantt} T'=(I-M')^{-1}\cdot S'.
\end{equation}

Then, the total amount of trade volume reduction in the entire
network is defined as $C_i$,
\begin{equation}
\label{eq.ci} C_i=(1,1,\cdot\cdot\cdot,1)\cdot(T-T').
\end{equation}

To ease our calculation, we always use the following equation
\begin{equation}
\label{eqn.cir}
C_i=\sum_{k=1}^N{\sum_{j=1}^N{S_j}\frac{{u_{ji}u_{ik}}}{u_{ii}}},
\end{equation}
where $U=(I-M)^{-1}$. It can be proved that Equation \ref{eqn.cir}
equals Equation \ref{eq.ci} (see SI).

\subsection*{Network Allometry}

Allometric scaling is a universal pattern of transportation networks
including rivers and vascular networks. Previous studies on network
allometry can only be applied to directed trees, in which $T_i$ is
the total number of nodes in the sub-tree rooted from $i$ and $C_i$
is the summation of all $T_i$s in the sub-tree rooted from
$i$\cite{Banavar_size_1999,Garlaschelli_universal_2003}.

The allometric exponents for trees are bounded between $1$ and $2$.
The minimum exponent can be obtained by a star-liked network, in
which all links are from the root to other nodes, whereas the
maximum exponent is obtained by a chain as shown in Figure
\ref{fig.starchain}. These two special trees stand for two extremes.
The star-liked tree is flat because every node except the root is
equivalent. However, the chain-liked tree is hierarchical because
the nodes at the upper level dominate the other nodes at the lower
level.

The allometric exponents can be computed for any tree. If the
exponent is smaller, then it is more star-like. Otherwise, it is
more chain-like.

Our previous works extended these studies to general flow
networks\cite{zhang_scaling_2010,zhang_allometry_2013}. By
incorporating the energy flow analysis technique, we can assign an
exponent for any flow network without edge cutting. By calculating
the $T_i$ and $C_i$ for each node $i$ according to the method
introduced in the previous subsection, we can obtain the allometric
exponent by estimating the slope of $\log T_i$ versus $\log C_i$.

The exponent $\eta$ can be viewed as the level of hierarchicality of
the flow structure because the relative speed of $C_i$, can increase
faster than $T_i$ in a network with a large exponent. As the
extension of the allometry of spanning trees, the flow network is
more like a chain if its exponent is large. Therefore, some long
flow chains can be revealed in these networks.

We distinguish networks as hierarchical ($\eta>1$), neutral ($\eta
\approx 1$) and flat ($\eta<1$) using the exponent.


\section*{Supporting Information Legends}
Table S1. The dataset form in UN dataset. Table S2. The trade data
in OECD dataset. Table S3. The value added data in OECD dataset.
Table S4. The result of   computed according to (4) and (5). Table
S5. Exponents of Leamer Classification Standard. Table S6. The top
ten Ci of different products in UN dataset. Table S7. Top ten
countries of different industries in the OECD Dataset. Figure S1.
Balanced value flow of one country. Figure S3. The relationship
between   and the mean proportion of foreign value added


\begin{thebibliography}{10}
\providecommand{\url}[1]{\texttt{#1}}
\providecommand{\urlprefix}{URL } \expandafter\ifx\csname
urlstyle\endcsname\relax
  \providecommand{\doi}[1]{doi:\discretionary{}{}{}#1}\else
  \providecommand{\doi}{doi:\discretionary{}{}{}\begingroup
  \urlstyle{rm}\Url}\fi
\providecommand{\bibAnnoteFile}[1]{%
  \IfFileExists{#1}{\begin{quotation}\noindent\textsc{Key:} #1\\
  \textsc{Annotation:}\ \input{#1}\end{quotation}}{}}
\providecommand{\bibAnnote}[2]{%
  \begin{quotation}\noindent\textsc{Key:} #1\\
  \textsc{Annotation:}\ #2\end{quotation}}
\providecommand{\eprint}[2][]{\url{#2}}

\bibitem{UNCTAD_global_2013}
{UNCTAD} (2013) Global value chains and development.
\newblock
  \urlprefix\url{http://unctad.org/en/PublicationsLibrary/diae2013d1\_en.pdf}.
\input{UNCTAD_global_2013}

\bibitem{Gibbon_Governing_2008}
Gibbon P, Bair J, Ponte S (2008) Governing global value chains: an
  introduction.
\newblock Economy and Society 37: 315--338.
\input{Gibbon_Governing_2008}

\bibitem{Gereffi_governance_2005}
Gereffi G, Humphrey J, Sturgeon T (2005) The governance of global
value chains.
\newblock Review of International Political Economy 12: 78--104.
\input{Gereffi_governance_2005}

\bibitem{Kotha_managing_2013}
Kotha S, Srikanth K (2013) Managing a global partnership model:
Lessons from
  the boeing 787 {'Dreamliner'} program.
\newblock Global Strategy Journal 3: 41--66.
\input{Kotha_managing_2013}

\bibitem{Rainnie_global_2013}
Rainnie A, Herod A, {McGrath-Champ} S (2013) Global production
networks, labour
  and small firms.
\newblock Capital \& Class 37: 177--195.
\input{Rainnie_global_2013}

\bibitem{Tukker_global_2013}
Tukker A, Dietzenbacher E (2013) Global multiregional input-output
frameworks:
  An introduction and outlook.
\newblock Economic Systems Research 25: 1--19.
\input{Tukker_global_2013}

\bibitem{Lenzen_building_2013}
Lenzen M, Moran D, Kanemoto K, Geschke A (2013) Building eora: A
global
  multi-region input-output database at high country and sector resolution.
\newblock Economic Systems Research 25: 20--49.
\input{Lenzen_building_2013}

\bibitem{Koopman_give_2010}
Koopman R, Powers W, Wang Z, Wei SJ (2010) Give credit where credit
is due:
  Tracing value added in global production chains.
\newblock Working Paper 16426, National Bureau of Economic Research.
\newblock \urlprefix\url{http://www.nber.org/papers/w16426}.
\input{Koopman_give_2010}

\bibitem{leontief_input-output_1966}
Leontief W (1966) Input-Output Economics.
\newblock Oxford University Press.
\input{leontief_input-output_1966}

\bibitem{Miller_input-output_2009}
Miller RE, Blair PD (2009) Input-output analysis: foundations and
extensions.
\newblock Cambridge [England]; New York: Cambridge University Press.
\input{Miller_input-output_2009}

\bibitem{Raa_economics_2005}
Raa Tt (2005) The economics of input-output analysis.
\newblock Cambridge: Cambridge University Press.
\input{Raa_economics_2005}

\bibitem{feenstra_world_2005}
Feenstra RC, Lipsey RE, Deng H, Ma AC, Mo H (2005) World trade
flows:
  1962-2000.
\newblock Working Paper 11040, National Bureau of Economic Research.
\newblock \urlprefix\url{http://www.nber.org/papers/w11040}.
\input{feenstra_world_2005}

\bibitem{zhu_compilation_2011}
Zhu S, Yamano N, Cimper A (2011) Compilation of bilateral trade
database by
  industry and end-use category.
\newblock {OECD} science, technology and industry working papers, Organisation
  for Economic Co-operation and Development, Paris.
\newblock
  \urlprefix\url{http://www.oecd-ilibrary.org/content/workingpaper/5k9h6vx2z07%
f-en}.
\input{zhu_compilation_2011}

\bibitem{Hidalgo_product_2007}
Hidalgo CA, Klinger B, Barabási AL, Hausmann R (2007) The product
space
  conditions the development of nations.
\newblock Science 317: 482--487.
\input{Hidalgo_product_2007}

\bibitem{Hidalgo_building_2009}
Hidalgo CA, Hausmann R (2009) The building blocks of economic
complexity.
\newblock Proceedings of the National Academy of Sciences 106: 10570--10575.
\input{Hidalgo_building_2009}

\bibitem{Fagiolo_world-trade_2009}
Fagiolo G, Reyes J, Schiavo S (2009) World-trade web: Topological
properties,
  dynamics, and evolution.
\newblock Physical Review E 79: 036115.
\input{Fagiolo_world-trade_2009}

\bibitem{garlaschelli_fitness-dependent_2004}
Garlaschelli D, Loffredo MI (2004) Fitness-dependent topological
properties of
  the world trade web.
\newblock Physical Review Letters 93: 188701.
\input{garlaschelli_fitness-dependent_2004}

\bibitem{Barigozzi_multinetwork_2010}
Barigozzi M, Fagiolo G, Garlaschelli D (2010) Multinetwork of
international
  trade: A commodity-specific analysis.
\newblock Physical Review E 81: 046104.
\input{Barigozzi_multinetwork_2010}

\bibitem{West_general_1997}
West G, Brown JH, Enquist BJ (1997) A general model for the origin
of
  allometric scaling laws in biology.
\newblock Science 276: 122--126.
\input{West_general_1997}

\bibitem{Banavar_size_1999}
Banavar J, Maritan A, Rinaldo A (1999) Size and form in efficient
  transportation networks.
\newblock Nature 399: 130--132.
\input{Banavar_size_1999}

\bibitem{Garlaschelli_universal_2003}
Garlaschelli D, Caldarelli G, Pietronero L (2003) Universal scaling
relations
  in food webs.
\newblock Nature 423: 165--168.
\input{Garlaschelli_universal_2003}

\bibitem{zhang_scaling_2010}
Zhang J, Guo L (2010) Scaling behaviors of weighted food webs as
energy
  transportation networks.
\newblock Journal of Theoretical Biology 264: 760--770.
\input{zhang_scaling_2010}

\bibitem{duan_universal_2007}
Duan W (2007) Universal scaling behavior in weighted trade networks.
\newblock European Physics Journal B 59: 271--276.
\input{duan_universal_2007}

\bibitem{Herrada_universal_2008}
Herrada EA, Tessone CJ, Klemm K, Eguiluz VM, Hernandez-Garcia E,
et~al. (2008)
  Universal scaling in the branching of the tree of life.
\newblock Plo{S ONE} 3: e2757.
\input{Herrada_universal_2008}

\bibitem{zhang_allometry_2013}
Zhang J, Wu L (2013) Allometry and dissipation of ecological flow
networks.
\newblock {PLoS} {ONE} 8: e72525.
\input{zhang_allometry_2013}

\bibitem{Foti_nonparametric_2011}
Foti NJ, Hughes JM, Rockmore DN (2011) Nonparametric sparsification
of complex
  multiscale networks.
\newblock {PLoS} {ONE} 6: e16431.
\input{Foti_nonparametric_2011}

\bibitem{Hausmann_what_2007}
Hausmann R, Hwang J, Rodrik D (2007) What you export matters.
\newblock Journal of Economic Growth 12: 1--25.
\input{Hausmann_what_2007}

\bibitem{cella_input-output_1984}
Cella G (1984) The input-output measurement of interindustry
linkages.
\newblock Oxford Bulletin of Economics and Statistics 46: 73�4.
\input{cella_input-output_1984}

\bibitem{song_linkage_2006}
Song Y, Liu C, Langston C (2006) Linkage measures of the
construction sector
  using the hypothetical extraction method.
\newblock Construction Management and Economics 24: 579--589.
\input{song_linkage_2006}

\end{thebibliography}
\end{document}